\documentclass[10pt]{revtex4}
\usepackage[pdftex]{graphicx}
\usepackage{epstopdf}
\usepackage{amssymb}
\usepackage{latexsym}
\usepackage{epsfig}
\usepackage{float}
\usepackage{graphicx,epsfig,color }
\usepackage{graphicx}
\usepackage{subfigure}
\usepackage{amsthm}
\usepackage{amsmath}

\usepackage{hyperref}
\begin{document}
\title{Joule-Thomson expansion of the quasitopological black holes}
\author{Behrouz Mirza\footnote{b.mirza@iut.ac.ir}, Fatemeh Naeimipour\footnote{sara.naeimipour1367@gmail.com}, Masoumeh Tavakoli\footnote{tavakoli.phy@gmail.com}}
\address{Department of Physics, Isfahan University of Technology, Isfahan 84156-83111, Iran}

\begin{abstract}
In this paper, we investigate the thermal stability and Joule-Thomson expansion of some new qusitopological black hole solutions. We first study the higher-dimensional static quasitopological black hole solutions in the presence of Born-Infeld, exponential and logarithmic nonlinear electrodynamics. The stable regions of these solutions are independent of the types of the nonlinear electrodynamics. The solutions with the horizons relating to the positive constant curvature, $k=+1$, have a larger region in thermal stability, if we choose positive quasitopological coefficients, $\mu_{i}>0$. We also have a review on the power Maxwell quasitopological black hole. Then, we obtain the five-dimensional Yang-Mills quasitopological black hole solution and compare with the quasitopological Maxwell solution. For large values of the electric charge, $q$, and the Yang-Mills charge, $e$, we showed that the stable range of the Maxwell quasitopological black hole is larger than the Yang-Mills one. 
This is while thermal stability for small charges has the same behavior for these black holes. In the following, we obtain the thermodynamic quantities for these solutions and then study the Joule-Thomson expansion.
  We consider the temperature changes in an isenthalpy process during this expansion. The obtained results show that the inversion curves can divide the isenthalpic ones into two parts in the inversion pressure, $P_{i}$. For $P<P_{i}$, a cooling phenomena with positive slope happens in $T-P$ diagram, while there is a heating process with negative slope for $P>P_{i}$. As the values of the  nonlinear parameter, $\beta$, the electric and Yang-Mills charges decrease, the temperature goes to zero with a small slope and so the heating phenomena happens slowly.
\end{abstract}


\maketitle

\section{Introduction}
Black holes as thermodynamic systems has been one of the most interesting subjects in theoretical physics for several decades. The main motivation for study thermodynamics of AdS black holes 
originates from the AdS/CFT correspondence. In this correspondence, the dynamics of the quantum field theory in $(d-1)$-dimensions is related to the dynamics of an AdS black hole in $d$-dimensions \cite{2}. The first-order phase transition of the AdS Schwarzschild black holes was studied by Hawking and Page \cite{4} and stimulated many physicists to study the phase structure of black holes. For exapmle, the thermodynamic phase structures of the charged AdS black holes have been studied in Refs.\cite{30,32}. In this paper, the cosmological constant has been considered as the dynamical pressure. P-V diagram of charged AdS black holes has the same behavior as the van der Waals liquid-gas phase transition.\\
Joule-Thomson expansion is an other thermodynamic issue which has attracted many people. For the van der Waals gases, the Joule-Thomson expansion is an isenthalpy process in which we can probe the temperature changes as the gas expands from the high pressure to the low one, through porous plugs. In fact, the Joule-Thomson expansion is a tool to know whether a cooling or heating process is happening for a gas. The zero value of the Joule-Thomson coefficient is called the inversion point in which the two cooling to heating processes intersect. Considering the mass of a black hole as an enthalpy \cite{115}, the Joule-Thomson expansion of charged AdS black holes was studied for the first time in Ref.\cite{okcu1}. The Joule-Thomson expansion were studied in many papers \cite{39,40,42,43Almeida,44,46,41,Rostami}. Joule-Thomson expansion of the higher dimensional charged AdS and Gauss-Bonnet AdS black holes have been broadly probed in respectively Refs. \cite{39,40}. In rainbow gravity, Joule-Thomson expansion of the Charged AdS black holes has also been studied in Ref. \cite{Jrainbow}. In Refs. \cite{Guo,Li}, the Joule-Thomson expansion of the usual and regular (Bardeen)-AdS black holes have been investigated. Joule-Thomson expansion of Born-Infeld AdS black holes has been investigated in Ref. \cite{JoulBI}. Born-Infeld nonlinear electrodynamics was introduced by Born and Infeld with the main aim to remove the divergence of the electric field \cite{Born}. Other types of the nonlinear electrodynamics such as logarithmic nonlinear(LN) and exponential nonlinear (EN) were introduced in Refs.\cite{Soleng} and \cite{Hen00}. The main purpose of this paper is to obtain the Joule-Thomson expansion for the nonlinear quasitopological black holes.\\
Quasitopological gravity is a higher curvature modified theory in $d$ dimensions. This gravity has some advantages which has attracted us to investigate. Based on the AdS/CFT correspondence, this gravity can provide a one-to-one duality between the central charges of the conformal field theory and the parameters in the gravitational side \cite{Lemos,Cai,Mann}. Also, Einstein's gravity is a low energy limit of the string theory which predicts higher dimensions. As the terms of the quasitopological gravity are not true topological invariants, so they can generate nontrivial gravitational terms in lower dimensions. This is the benefit of this gravity to the other modified gravities such as Lovelock. This gravity can also provide a dual CFT which respects the causality \cite{Myers}. Quasitopological black holes and $P-V$ criticality behaviour of them have studied in Refs.\cite{Mann2012,Man1}.  
Thermodynamics of AdS black hole and holography  in generalized quasi-topological gravity was investigate in Refs. \cite{Mir1,Mir2}.
\\
We also obtained the five-dimensional Yang-Mills (YM) quasitopological black hole solutions and compared them with the ones in quasitopological Maxwell theory. The YM theory is one of the attractive non-abelian gauge theories that comes from the low energy limit of the string theory models spectrum.   
Non-abelian gauge fields beside the gravitational ones can be an effective subject in the physical phenomena to the results of superstring models. The analytic black hole solution of the Einstein-YM (EYM) was firstly developed by Yasskin in Ref.\cite{yass1}. Black hole solutions in the presence of nonabelian Yang-Miils theory have been obtained in Refs. \cite{18ym, 19ym, 21ym, 22ym}.  Black holes in a non-abelian Born-Infeld theory and supersymmetric EYM theories were studied in \cite{6ym, 7ym}, respectively. Using the Wu-Yang ansatz \cite{Wu}, black hole solutions of the various gravities coupled to YM field have been explored in Refs.\cite{2ym, 3ym,4ym, 5ym,Bost,Deh,24ym}. It is interesting to look at the Joule-Thomson expansion of the Yang-Mills quasitopological solutions. \\
This paper is arranged as follows: We start with the quasitopological gravity and the nonlinear electrodynamic theory in Sec. \ref{Field} and find the related static solutions. Then, we obtain the thermodynamic quantities and study the thermal stability of the related solutions in respectively Secs. \ref{thermo} and \ref{stab}. 
We also probe the Joule-Thomson expansion of the Power Maxwell quasitopological black hole in Sec.\ref{Joule2}. In Sec.\ref{Joule3}, we obtain the solution of the Yang-Mills black hole in the presence of the quasitopological gravity and then probe th thermal stability and Joule-Thomson expansion for this black hole. At last, we have a conclusion of the whole paper in Sec. \ref{con}. 
\section{The static solutions of the $(n+1)$-dimensional nonlinear quasitopological gravity}\label{Field}
The main structure of the $(n+1)$-dimensional quasitopological gravity starts from the action \cite{Baz1,Ghana1,Naeimi3} 
\begin{equation}\label{Act1}
I_{\rm{bulk}}=\frac{1}{16\pi}\int{d^{n+1}x\sqrt{-g}\big\{-2\Lambda+R+\hat{\mu}_{2} {\mathcal L}_2+\hat{\mu}_{3} {\mathcal L}_3+\hat{\mu}_{4}{\mathcal L}_4+\mathcal{L}(F)\big\}},
\end{equation}
where $\mathcal{L}(F)$ is the matter source
that for the nonlinear electrodynamics is considered as follows:
\begin{eqnarray}\label{non1}
\mathcal{L}(F)=\left\{
\begin{array}{ll}
$$4\beta^2[1-\sqrt{1+\frac{F^2}{2\beta^2}}]$$,\quad\quad\quad \quad\quad  \ {BI}\quad &  \\ \\
$$4\beta^2[\mathrm{exp}(-\frac{F^2}{4\beta^2})-1]$$,\quad\quad\quad \quad\,\,\, \ {EN}\quad &  \\ \\
$$-8\beta^2 \mathrm{ln}[1+\frac{F^2}{8\beta^2}]$$,\quad\quad\quad\quad\quad\quad\,  \ {LN}\quad &
\end{array}
\right.
\end{eqnarray}
where $BI$, $EN$ and $LN$ are  the abbreviation of the Born-Infeld, exponential and logarithmic forms, respectively \cite{Born,Soleng,Hen00}.

We define $F^2=F_{\mu\nu}F^{\mu\nu}$, where the electromagnetic field tensor is described as $F_{\mu\nu}=\partial_{\mu}A_{\nu}-\partial_{\nu}A_{\mu}$ with $A_{\mu}$ as the vector potential. The Lagrangians ${\mathcal L}_2$, ${\mathcal L}_3$ and ${{\mathcal L}_4}$ are respectively referred to the second-order Lovelock (Gauss-Bonnet), cubic and quartic quasitopological gravity with the constant coefficients $\hat{\mu}_2$, $\hat{\mu}_3$ and $\hat{\mu}_4$. It should be noted that a hat($\,\hat{}\,$) is not the sign of an operator. ${\mathcal L}_2$, ${\mathcal L}_3$ and ${{\mathcal L}_4}$ are defined as \cite {Baz1}
\begin{eqnarray}
{\mathcal L}_2&=& R_{abcd}R^{abcd}-4R_{ab}R^{ab}+R^2,
\end{eqnarray}
\begin{eqnarray}\label{quasi3}
{{\mathcal L}_3}&=&
R_a{{}^c{{}_b{{}^d}}}R_c{{}^e{{}_d{{}^f}}}R_e{{}^a{{}_f{{}^b}}}+\frac{1}{(2n-1)(n-3)} \bigg(\frac{3(3n-5)}{8}R_{abcd}R^{abcd}R-3(n-1)R_{abcd}R^{abc}{{}_e}R^{de}\nonumber\\
&&+3(n+1)R_{abcd}R^{ac}R^{bd}+6(n-1)R_a{{}^b}R_b{{}^c}R_{c}{{}^a}-\frac{3(3n-1)}{2}R_a{{}^b}R_b{{}^a}R +\frac{3(n+1)}{8}R^3\bigg),
\end{eqnarray}
\begin{eqnarray}\label{quasi4}
{\mathcal{L}_4}&=& c_{1}R_{abcd}R^{cdef}R^{hg}{{}_{ef}}R_{hg}{{}^{ab}}+c_{2}R_{abcd}R^{abcd}R_{ef}{{}^{ef}}+c_{3}RR_{ab}R^{ac}R_c{{}^b}+c_{4}(R_{abcd}R^{abcd})^2\nonumber\\
&&+c_{5}R_{ab}R^{ac}R_{cd}R^{db}+c_{6}RR_{abcd}R^{ac}R^{db}+c_{7}R_{abcd}R^{ac}R^{be}R^d{{}_e}+c_{8}R_{abcd}R^{acef}R^b{{}_e}R^d{{}_f}\nonumber\\
&&+c_{9}R_{abcd}R^{ac}R_{ef}R^{bedf}+c_{10}R^4+c_{11}R^2 R_{abcd}R^{abcd}+c_{12}R^2 R_{ab}R^{ab}\nonumber\\
&&+c_{13}R_{abcd}R^{abef}R_{ef}{{}^c{{}_g}}R^{dg}+c_{14}R_{abcd}R^{aecf}R_{gehf}R^{gbhd},
\end{eqnarray}
where the coefficients $c_{i}$'s are written in the appendix \eqref{app1}. To find a handle of the static topological solutions, we use the metric
\begin{eqnarray}\label{metric1}
ds^2=-f(r) dt^2+\frac{d r^2}{f(r)}+r^2 d\Omega_{k,n-1}^2,
\end{eqnarray}
where $d\Omega_{k,n-1}^2$ represents the line element of a $(n-1)$-dimensional
hypersurface $\Sigma$ with the constant curvature $k=1,0,-1$ as the spherical, flat and hyperbolic geometries, respectively. By varying the action \eqref{Act1} with respect to $A_{\mu}$ and solving the related equation, so the electromagnetic field tensor is obtained as 
\begin{eqnarray}\label{Ftr}
F_{tr}=\left\{
\begin{array}{ll}
$$\frac{q}{r^{n-1}}(\sqrt{1+\eta})^{-1}$$,\quad \quad\quad\quad \quad\quad\quad \ {BI}\quad &  \\ \\
$$\beta\sqrt{L_{W}(\eta)}$$,\quad \quad\quad\quad \quad\quad\quad\quad\quad\,\,  \ {EN}\quad &  \\ \\
$$\frac{2q}{r^{n-1}}(1+\sqrt{1+\eta})^{-1}$$,\quad\quad\quad\quad\quad\,\,  \ {LN}\quad &
\end{array}
\right.
\end{eqnarray}
where $\eta=\frac{q^2}{\beta^2 r^{2n-2}}$ and $q$ is an integration constant. If we vary the action \eqref{Act1} with respect to $g_{\mu\nu}$ and redefine the quasitopological gravity coefficients as
\begin{eqnarray}
\mu_{2}=(n-2)(n-3) \hat{\mu}_{2},
\end{eqnarray}
\begin{eqnarray}
\mu_{3}=\frac{(n-2)(n-5)(3n^2-9n+4)}{8(2n-1)}\hat{\mu}_{3},
\end{eqnarray}
\begin{eqnarray}
\mu_{4}=n(n-1)(n-3)(n-7)(n-2)^2(n^5-15n^4+72n^3-156n^2+150n-42)\hat{\mu}_{4},
\end{eqnarray}
so the gravitational field equation is gained as 
\begin{eqnarray}\label{Eq}
\mu_{4} \Psi^4+\mu
_{3} \Psi^3+\mu
_{2} \Psi^2+\Psi+\xi=0,
\end{eqnarray}
where we have the definitions $\Psi(r)=[k-f(r)]/r^2$ and  
\begin{eqnarray}\label{kappa1}
\xi&=&-\frac{2\Lambda}{n(n-1)}-\frac{m}{r^n}+\nonumber\\
&&\left\{
\begin{array}{ll}
$$\frac{4\beta^2}{n(n-1)}(1-{}_2 F_{1}([-\frac{1}{2},-\frac{n}{2(n-1)}]\,,[\frac{n-2}{2(n-1)}]\,,-\eta))$$,\quad \quad\quad\quad\quad\quad \quad\quad\quad\quad\quad\quad\quad\quad\quad\quad\quad  \ {BI}\quad &  \\ \\
$$-\frac{4\beta^2}{n(n-1)}+\frac{4(n-1)\beta q}{n(n-2)r^n}(\frac{q}{\beta})^{\frac{1}{n-1}}(L_{W}(\eta))^{\frac{n-2}{2(n-1)}}\times{}_2 F_{1}([\frac{n-2}{2(n-1)}]\,,[\frac{3n-4}{2(n-1)}]\,,-\frac{1}{2(n-1)}L_{W}(\eta))\\-\frac{4\beta q}{(n-1)r^{n-1}}[L_{W}(\eta)]^{\frac{1}{2}}\times[1-\frac{1}{n}(L_{W}(\eta))^{-1}]$$,\quad \quad\quad\quad\quad\quad \quad\quad\quad\quad\quad\quad\quad\quad\quad \quad\quad\quad\,\,   \ {EN}\quad &  \\ \\
$$\frac{8(2n-1)}{n^2(n-1)}\beta^2[1-\sqrt{1+\eta}]+\frac{8(n-1)q^2}{n^2 (n-2)r^{2n-2}}{}_2 F_{1}([\frac{n-2}{2(n-1)},\frac{1}{2}]\,,[\frac{3n-4}{2(n-1)}]\,,-\eta)\\
-\frac{8}{n(n-1)}\beta^2 \mathrm {ln}[\frac{2\sqrt{1+\eta}-2}{\eta}]$$,\quad\quad\quad\quad\quad\quad\quad\quad\quad\quad\quad\quad\quad\quad\quad\quad\quad\quad \quad\quad\quad\quad \quad\quad \quad\quad\,\,\,   \ {LN}\quad &
\end{array}
\right.
\end{eqnarray}
where $m$ is an integration constant relating to the mass of the black hole. In the above relation, $L_W(x)$ and ${}_2 F_{1}([a,b],[c],z)$ are respectively the Lambert and hypergeometric functions.
We get to the solution for equation (11)
\begin{eqnarray}\label{func1}
f(r)=k-r^2\times\left\{
\begin{array}{ll}
$$-\frac{\mu_{3}}{4\mu_{4}}+\frac{-W+\sqrt{-(3A+2y-\frac{2B}{W})}}{2}$$,\quad \quad\quad\quad \  \ {\mu_{4}>0,}\quad &  \\ \\
$$-\frac{\mu_{3}}{4\mu_{4}}+\frac{W-\sqrt{-(3A+2y+\frac{2B}{W})}}{2}$$, \quad \quad\quad\quad\quad\,\,{\mu_{4}<0,}\quad &
\end{array}
\right.
\end{eqnarray} 
where
for brevity reason,
 we have described $W$, $A$, $y$ and $B$ in the appendix \eqref{app2}.
\section{Thermodynamic behavior of the $(n+1)$-dimensional static nonlinear quasitopological black hole}\label{thermo}
Via the AdS/CFT correspondence, the thermodynamic behaviors of an AdS black hole can provide a set of knowledge for a certain dual conformal field theory (CFT). So, in this section, we are eager to obtain the thermodynamic quantities of the static nonlinear quasitopological black hole. Using the subtraction method \cite{Brown}
, the mass of this black hole is gained as
\begin{eqnarray}\label{mass}
M=\frac{(n-1)}{16\pi}m,
\end{eqnarray}
where $m$ can be obtained form Eq.\eqref{Eq} by the fact that $f(r_{+})=0$. The electric charge of the black hole can be determined from the Gauss law
\begin{eqnarray}
Q=\frac{1}{4\pi}\int\,F_{tr}r^{n-1} d\Omega_{k}=\frac{q}{4\pi},
\end{eqnarray}
and the electric potential $U$ is defined by the formula $U=A_{\nu}\chi^{\nu}|_{\infty}-A_{\nu}\chi^{\nu}|_{r=r_{+}}$, where $\chi^{\nu}$ is the killing vector. So, we can get to the electric potential as follow
\begin{eqnarray}\label{potential}
U=\left\{
\begin{array}{ll}
$$\frac{q}{(n-2)r_{+}^{n-2}}{}_{2}F_{1}([\frac{1}{2},\frac{n-2}{2(n-1)}]\,,[\frac{3n-4}{2(n-1)}]\,,-\eta_{+})$$,\quad \quad\quad\quad \quad\quad\quad\quad\quad\quad \quad\quad\quad \quad\quad\quad \quad\quad\quad \quad\quad\quad \quad\,\,\,  \ {BI}\quad &  \\ \\
$$\frac{n-1}{n-2}\beta\big(\frac{q}{\beta}\big)^{\frac{1}{n-1}}\big(L_{W}(\eta_{+})\big)^{\frac{n-2}{2(n-1)}} {}_2F_{1}\bigg(\big[\frac{n-2}{2(n-1)}\big]\,,\big[\frac{3n-4}{2(n-1)}\big]\,,-\frac{1}{2(n-1)}L_{W}(\eta_{+})\bigg)
-\beta r_{+} \sqrt{L_{W}(\eta_{+})}$$,\quad \quad \ {EN}\quad &  \\ \\
$$\frac{q}{(n-2)r_{+}^{n-2}}{}_{3}F_{2}([\frac{n-2}{2(n-1)},\frac{1}{2},1]\,,[\frac{3n-4}{2(n-1)},2]\,,-\eta_{+})$$,\quad\quad\quad\quad\quad\quad\quad\quad\quad\quad \quad\quad\quad \quad\quad\quad \quad\quad\quad \quad\quad\,  \ {LN}\quad &
\end{array}
\right.
\end{eqnarray}
where $\eta_{+}\equiv\eta(r=r_{+})$. The entropy \cite{Wald} and the Hawking temperature of the static nonlinear quasitopological black hole can be obtained by
\begin{eqnarray}\label{entropy}
S=\frac{n-1}{4}r_{+}^{n-1}\bigg(\frac{1}{n-1}+ \frac{2k\mu_{2}}{(n-3)r_{+}^2}+ \frac{3k^2\mu_{3}}{(n-5)r_{+}^4}+ \frac{4k^3\mu_{4}}{(n-7)r_{+}^6}\bigg),
\end{eqnarray}  
\begin{eqnarray}\label{Temp}
T&=&\frac{f^{'}(r_{+})}{4\pi}=\frac{1}{4\pi r_{+}(4\mu_{4}k^3+3\mu_{3}k^2r_{+}^2+2\mu_{2}k r_{+}^4+r_{+}^6)}\times [\mu_{4}k^4(n-8)+\mu_{3}k^3(n-6)r_{+}^2\nonumber\\
&&+\mu_{2}k^2(n-4)r_{+}^4+k(n-2)r_{+}^6-\frac{2\Lambda}{n-1} r_{+}^8+\left\{
\begin{array}{ll}
$$\frac{4\beta^2}{n-1} r_{+}^8(1-\sqrt{1+\eta_{+}})]$$,\quad \quad\quad\quad \quad\quad\quad\quad\quad\quad\quad\quad\,\, \ {BI}\quad &  \\ \\
$$-\frac{4\beta^2}{n-1} r_{+}^8\{1-(1-L_{W}(\eta_{+}))\mathrm{exp}(L_{W}(\eta_{+})/2)\}]$$,\quad \quad \ {EN}\quad &  \\ \\
$$\frac{8\beta^2 }{n-1}r_{+}^8[1-\sqrt{1+\eta_{+}}-\mathrm{ln}\big(\frac{2[\sqrt{1+\eta_{+}}-1]}{\eta_{+}}\big)]]$$.\quad\quad\quad\quad\,\,\, \ {LN}\quad &
\end{array}
\right.
\end{eqnarray}
If we consider the thermodynamic volume and pressure as bellow \cite{Dolan}
\begin{eqnarray}\label{Press}
&&V=\frac{r_{+}^n}{n},\nonumber\\
&&P=-\frac{\Lambda}{8\pi},
\end{eqnarray}
therefore, the first law in the extended phase space follows from the formula  
\begin{eqnarray}
dM=TdS+UdQ+VdP+Bd\beta+\Psi_{2}d\mu_{2}+\Psi_{3}d\mu_{3}+\Psi_{4}d\mu_{4},
\end{eqnarray}
where $B$ and $\Psi_{i}$'s,($i=1,2,3$) are denoted respectively as the potentials conjugate to the nonlinear parameter $\beta$ and couplings $\mu_{i}$'s, respectively. They are defined as follows
\begin{eqnarray}
&&B=\frac{\partial M}{\partial \beta}\,\,,\,\,\nonumber\\
&&\Psi_{2}=\frac{\partial M}{\partial \mu_{2}}=\frac{(n-1)k^2 r_{+}^{n-4}}{16\pi}-\frac{(n-1)k r_{+}^{n-3}}{2(n-3)}T\,\,\,,\,\,\,\nonumber\\
&&\Psi_{3}=\frac{\partial M}{\partial \mu_{3}}=\frac{(n-1)k^3 r_{+}^{n-6}}{16\pi}-\frac{3(n-1)k^2r_{+}^{n-5}}{4(n-5)}T\,\,\,,\,\,\,\nonumber\\
&&\Psi_{4}=\frac{\partial M}{\partial \mu_{4}}=\frac{(n-1)k^4 r_{+}^{n-8}}{16\pi}-\frac{(n-1)k^3r_{+}^{n-7}}{(n-7)}T,
\end{eqnarray}
that the relations $ T=\frac{\partial M}{\partial S}\,\,,\,\,U=\frac{\partial M}{\partial Q}\,\,,\,\,V=\frac{\partial M}{\partial P}$ are established.
In the extended phase space, we can write the Smarr-type formula of this black hole as
\begin{eqnarray}
M=\frac{1}{n-2}[(n-1)TS-2PV+(n-2)UQ-\beta B+2\mu_{2}\Psi_{2}+4\mu_{3}\Psi_{3}+6\mu_{4}\Psi_{4}].
\end{eqnarray}
If we determine the specific volume $v=\frac{4r_{+}}{n-1}$ and use the pressure \eqref{Press} in Eq.\eqref{Temp}, so we can specify the equation of state for the static nonlinear quasitopological black hole as $P(v,T)$. The critical points of the static nonlinear quasitopological black hole can be derived from the following conditions 
\begin{eqnarray}\label{cons}
\frac{\partial P}{\partial v}|_{v_{C}}=0\,\,\,,\,\,\,\frac{\partial ^2 P}{\partial v^2}|_{v_{C}}=0.
\end{eqnarray}
Critical behavior of the cubic quasitopological black hole has been investigated in Ref.\cite{Man1}. As the critical behavior of the quartic quasitopological black hole is similar to the cubic one, so we avoid repeating them here.\\
\section{Thermal stability of the $(n+1)$-dimensional static nonlinear quasitopological black hole}\label{stab}
\begin{figure}
\centering
\subfigure[Temperature $T$]{\includegraphics[scale=0.27]{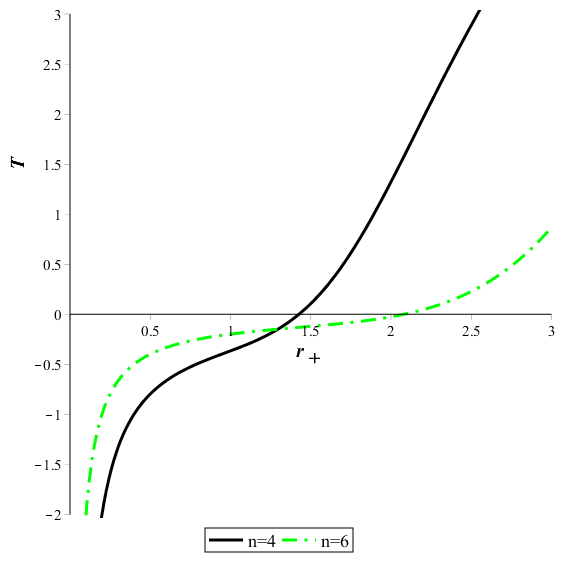}\label{Fig1a}}\hspace*{.2cm}
\subfigure[Heat capacity $C_{P}$]{\includegraphics[scale=0.27]{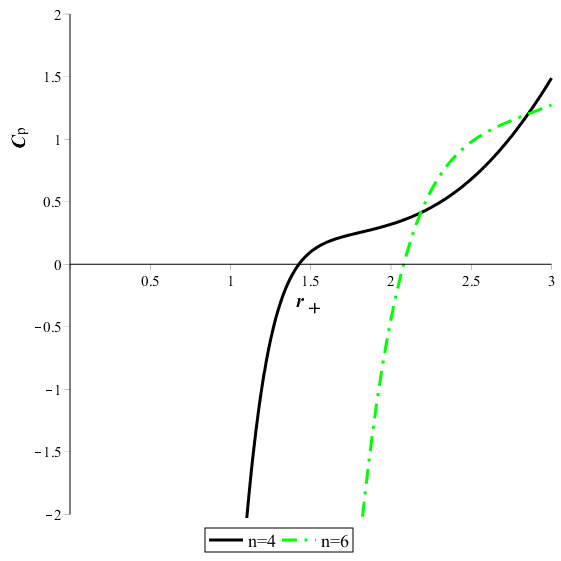}\label{Fig1b}}\caption{Thermal stability of the BI quasitopological black hole with respect to $r_{+}$ for different values of dimension $n$ with $\hat{\mu}_{2}=0.2$, $\hat{\mu}_{3}=0.1$, $\hat{\mu}_{4}=0.001$ $k=1$, $q=1$ and $\beta=6$.}\label{Fig1}
\end{figure}

\begin{figure}
\centering
\subfigure[Temperature $T$]{\includegraphics[scale=0.27]{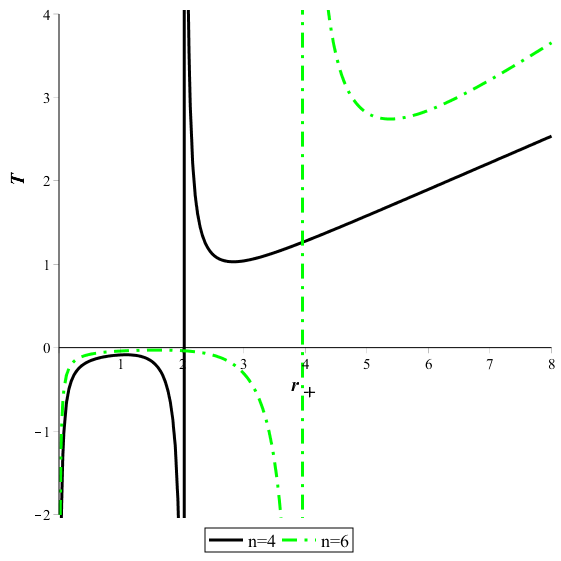}\label{Fig2a}}\hspace*{.2cm}
\subfigure[Heat capacity $C_{P}$]{\includegraphics[scale=0.27]{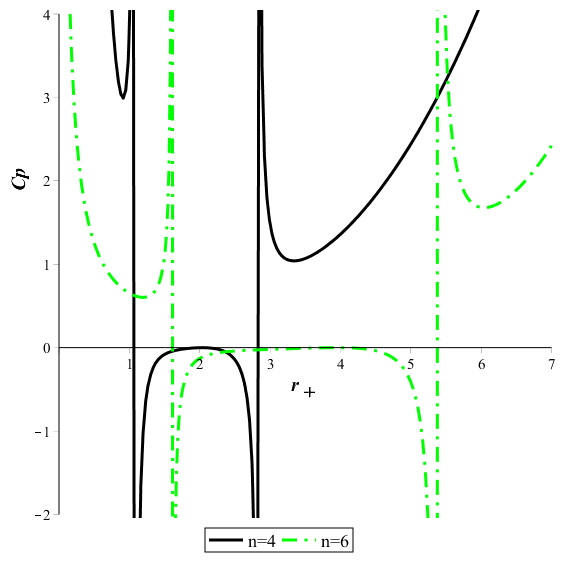}\label{Fig2b}}\caption{ Thermal stability of the BI quasitopological black hole with respect to $r_{+}$ for different values of dimension $n$ with $\hat{\mu}_{2}=0.2$, $\hat{\mu}_{3}=0.1$, $\hat{\mu}_{4}=-0.001$ $k=1$, $q=1$ and $\beta=6$.}\label{Fig2}
\end{figure}

In order to know where a black hole may exist physically or no, we should discuss about its thermal stability. To study the thermal stability of the static nonlinear quasitopological black hole, we define the heat capacity $C_{P}$ at the constant pressure as follow
\begin{eqnarray}\label{heatcap}
C_{P}=T\bigg(\frac{\partial S}{\partial T}\bigg)_{P}=T\frac{\big(\frac{\partial S}{\partial r_{+}}\big)_{P}}{\big(\frac{\partial T}{\partial r_{+}}\big)_{P}}.
\end{eqnarray}
The positive value of $C_{P}$ may lead to the thermal stability of the mentioned black hole, while the negative value shows the instability. We should note that the positive value of the temperature is a requirement in order to have physical solutions. To show the stability of the static nonlinear quasitopological black hole, we have plotted $C_{P}$ and $T$ for BI black hole in Figs.\ref{Fig1} and \ref{Fig2} with $k=1$ and $\hat{\mu}_{2}>0$ and $\hat{\mu}_{3}>0$. The obtained results show that the type of the nonlinear electrodynamics has a trivial effect on the thermal stability. 
Therefore, we have refused probing all of them and just included the stability of the BI theory.
In Fig.\ref{Fig1} with $\hat{\mu}_{4}>0$, we can see a $r_{+\mathrm{min}}$ for each $n=4$ and $6$ dimensions, that $C_{P}$ and $T$ are both positive for $r_{+}>r_{+\mathrm{min}}$. For $\hat{\mu}_{4}<0$ in Fig.\ref{Fig2b}, there are two $r_{+\mathrm{min}}$ and $r_{+\mathrm{max}}$ which $C_{P}$ is positive for both regions $r_{+}<r_{+\mathrm{min}}$ and $r_{+}>r_{+\mathrm{max}}$. 
According the temperature diagram in Fig.\ref{Fig2a}, a unit positive region for both $C_{P}$ and $T$ can be gained for just $r_{+}>r_{+\mathrm{max}}$. Comparing figures \ref{Fig1} and $\ref{Fig2}$ shows that for the same parameters, a black hole with $\hat{\mu}_{4}>0$ may have a larger region in thermal stability than the one with $\hat{\mu}_{4}<0$. We can also understand this result from Eqs. \eqref{entropy}, \eqref{Temp} and \eqref{heatcap}. The nonlinear quasitopological black holes with $k=1$ and positive $\hat{\mu}_{2}$ and $\hat{\mu}_{3}$ can have larger positive regions for $\partial S/\partial r_{+}$, $T$ and $C_{P}$, if we choose $\hat{\mu}_{4}>0$. \\
\section{Joule-Thomson expansion of the $(n+1)$-dimensional static nonlinear quasitopological black hole}\label{Joule1}
\begin{figure}
\centering
\subfigure[Joule-Thomson coefficient $\mu$]{\includegraphics[scale=0.27]{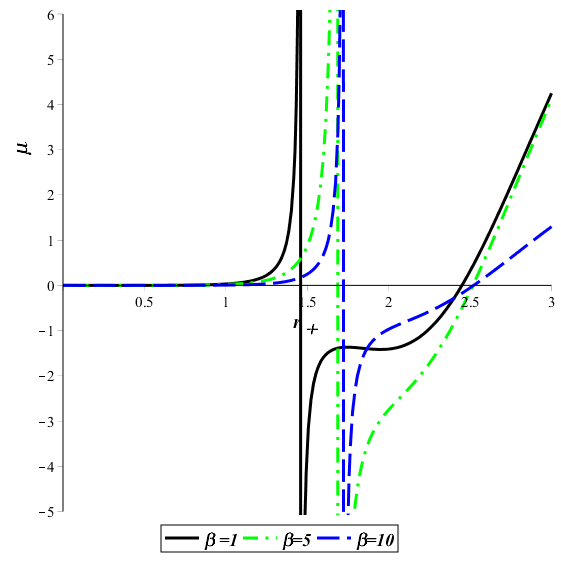}\label{Fig3a}}\hspace*{.2cm}
\subfigure[Temperature $T$]{\includegraphics[scale=0.27]{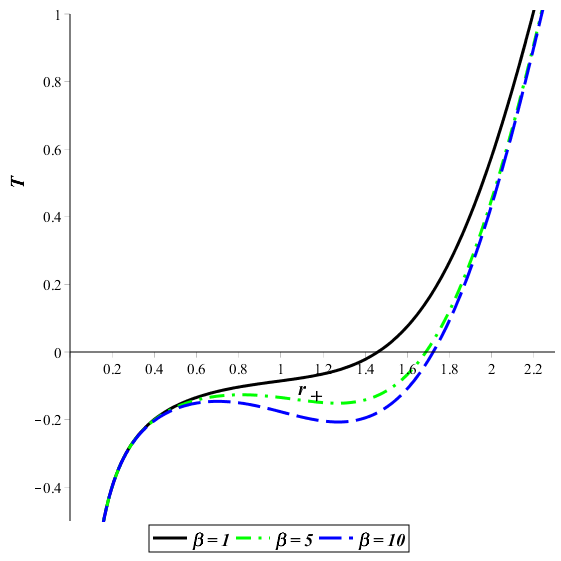}\label{Fig3b}}\caption{The Joule-Thomson coefficient $\mu$ and temperature $T$ of the BI quasitopological black hole with respect to $r_{+}$ for different values of the nonlinear parameter $\beta$ with $\hat{\mu}_{2}=0.1$, $\hat{\mu}_{3}=0.1$, $\hat{\mu}_{4}=0.002$, $k=1$, $Q=2$ and $n=4$.}\label{Fig3}
\end{figure}

\begin{figure}
\centering
\subfigure[$Q=3$, $\beta=10$]{\includegraphics[scale=0.27]{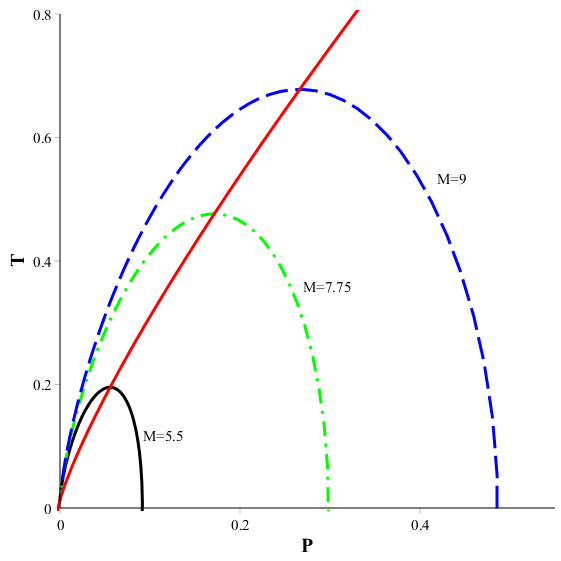}\label{Fig4a}}\hspace*{.2cm}
\subfigure[$Q=3$,$\beta=1$]{\includegraphics[scale=0.27]{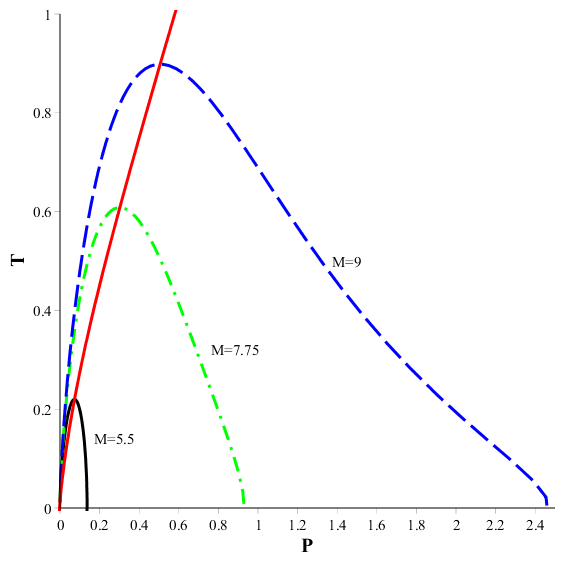}\label{Fig4b}}\hspace*{.2cm}
\subfigure[$Q=1$,$\beta=1$]{\includegraphics[scale=0.27]{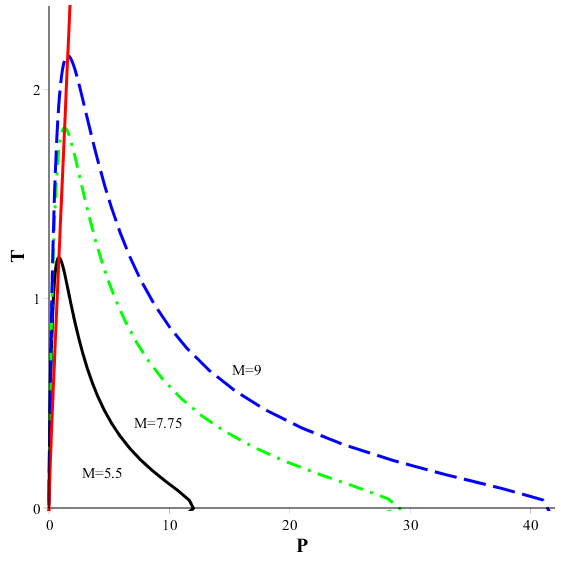}\label{Fig4c}}\caption{Isenthalpic curves and inversion curve of the BI quasitopological black hole with $\hat{\mu}_{2}=0.1$, $\hat{\mu}_{3}=0.1$, $\hat{\mu}_{4}=0.001$, $k=1$ and $n=4$.}\label{Fig4}
\end{figure}

\begin{figure}
\center
\includegraphics[scale=0.25]{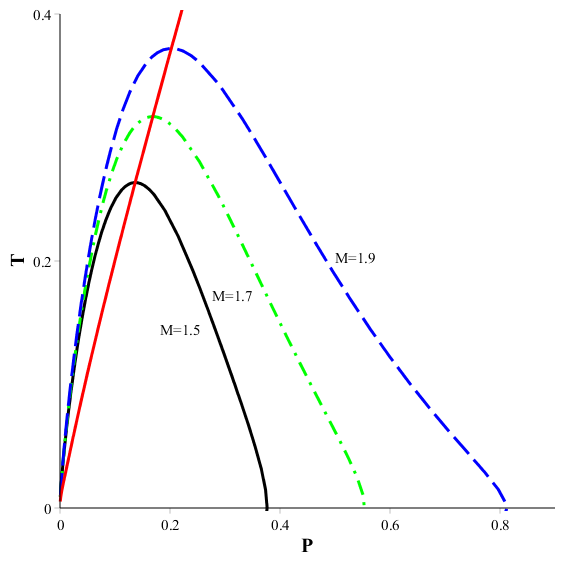}
\caption{\small{Isenthalpic curves and inversion curve of the BI quasitopological black hole with $\hat{\mu}_{2}=0.1$, $\hat{\mu}_{3}=0.1$, $\hat{\mu}_{4}=-0.001$, $k=-1$, $Q=1$, $n=4$ and $\beta=20$.} \label{Fig5}}
\end{figure}

In this section, we 
intend
to study the Joule-Thomson expansion of the obtained quasitopological black hole coupled to the nonlinear electrodynamics. In the classical thermodynamics, the Joule-Thomson expansion is an isenthalpy process in which we can probe the temperature changes as the gas expands from the high pressure to the low one
through porous plugs. The Joule-Thomson coefficient is obtained by the formula \cite{okcu1}
\begin{eqnarray}\label{formule1}
\mu=\bigg(\frac{\partial T}{\partial P}\bigg)_{H}=\frac{1}{C_{P}}\big[T\bigg(\frac{\partial V}{\partial T}\bigg)_{P}-V\big],
\end{eqnarray}
where the enthalpy of the system, $H$, is fixed. In the gas expansion, the pressure always decreases. So, when the value of the coefficient $\mu$ is positive during the expansion, it means that the temperature decreases and therefore it is called a cooling phenomenon. However, when $\mu$ is negative, the temperature increases and this is called a heating process. For $\mu=0$, we can obtain the inversion temperature in which the process of the temperature changes gets vice versa. It can be obtained by the formula
\begin{eqnarray}\label{invesion}
T_{i}=V\bigg(\frac{\partial T}{\partial V}\bigg)_{P}.
\end{eqnarray}
As a black hole behaves like a thermodynamic system, so we can consider the mass of a black hole as the enthalpy and probe the Joule-Thomson expansion for it. \\
Now, we would like to investigate the Joule-Thomson expansion of the higher-dimensional nonlinear quasitopological black hole and identify the region in which cooling or heating occurs. Therefore in Fig.\ref{Fig3}, we have plotted the Joule-Thomson coefficient $\mu$ versus $r_{+}$ for different values of $\beta$ and compared it with the temperature of the black hole. In Fig.\ref{Fig3a}, there is a $r_{+\mathrm{ext}}$ for each values of $\beta$ in which the coefficient $\mu$ diverges. This point is in accordance with $T=0$ in Fig. \ref{Fig3b} where there is an extreme black hole. So, we can get some knowledge about the extremal black hole by recognizing the infinite points of $\mu$. This figure also shows that by increasing the nonlinear parameter $\beta$, the value of $r_{+\mathrm{ext}}$ increases. For $r_{+}>r_{+\mathrm{ext}}$ in Fig.\ref{Fig3a}, there is also an inversion phenomenon in $r_{+\mathrm{inv}}$ in which the black hole goes from a heating process to a cooling one. For small $\beta$, inversion happens in a smaller $r_{+}$.\\
We have also plotted the isenthalpic curves and the inversion curve of the nonlinear quasitopological black hole for different values of $Q$ and $\beta$ in Fig.\ref{Fig4}. In each subfigure, we can see three isenthalpic curves with constant $M$ and the related inversion curve happening at the maximum value of the isenthalpic curves. We define the inversion temperature and pressure of each isenthalpic as $T_{i}$ and $P_{i}$. The inversion curve divides the isenthalpic curves in to two parts where for $P<P_{i}$, the slope of the isenthalpic curve is positive and so a cooling happens in the expansion. But, for $P>P_{i}$, the slope of isenthalpic curve is negative and so there is a heating for the black hole. For small values of parameter $\beta$ in Figs. \ref{Fig4b} and \ref{Fig4c}, the same behaviors are repeated, but for $P>P_{i}$, the temperature decreases to zero with a steeper slope. So, the heating process happens slowly. This is unlike the Einstein-Born-Infeld black hole for which the slope of the curve in the range $P>P_{i}$ is unchanged as $\beta$ increases \cite{JoulBI}. By decreasing the parameter $\beta$ in Fig.\ref{Fig4b} with respect to the one in Fig. \ref{Fig4a}, the extreme black hole will happen in a larger pressure. In Fig.\ref{Fig4c} with small electric charge, the heating happens very slowly and so the temperature gets to zero for a higher pressure value than the figure \ref{Fig4b} with larger charge. 
We have also checked out the Joule-Thomson expansion of the obtained black hole for $\hat{\mu}_{4}<0$ in Figure \ref{Fig5}. The result shows that we can face to isenthalpic curves with $\hat{\mu}_{4}<0$ just for the hyperbolic geometry, $k=-1$.  

\section{Joule-Thomson expansion of the power Maxwell quasitopological black hole}\label{Joule2}
\begin{figure}
\centering
\subfigure[$s=0.7$]{\includegraphics[scale=0.27]{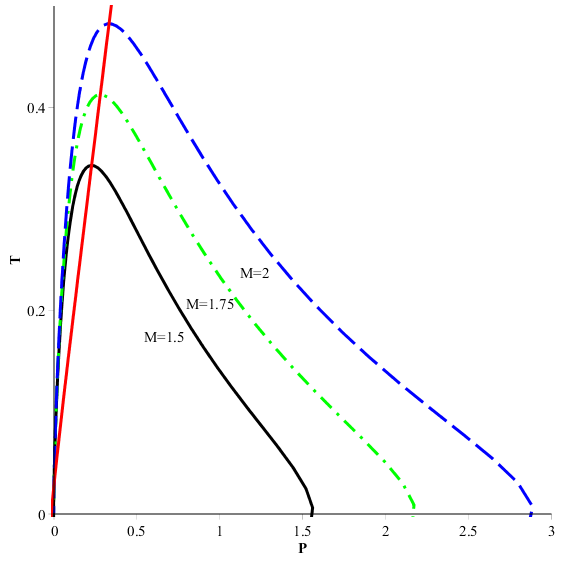}\label{fig6a}}\hspace*{.2cm}
\subfigure[$s=1.4$]{\includegraphics[scale=0.27]{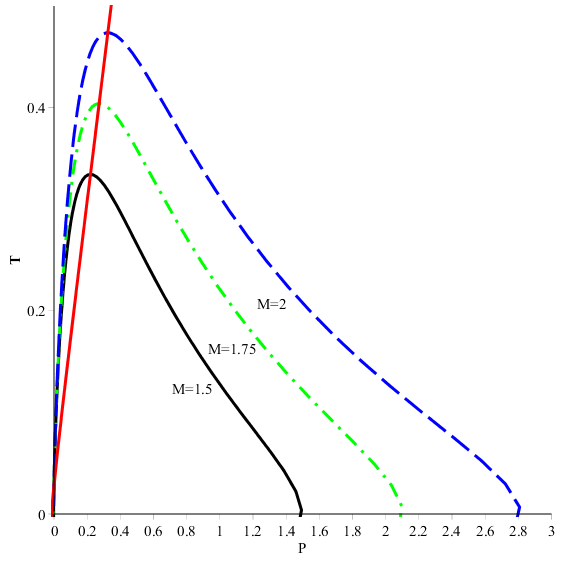}\label{fig6b}}\caption{Isenthalpic curves and inversion curve of the PM quasitopological black hole with $\hat{\mu}_{2}=0.1$, $\hat{\mu}_{3}=0.1$, $\hat{\mu}_{4}=0.001$, $k=1$, $n=4$ and $q=1$.} \label{Fig6}
\end{figure}

Power Maxwell is an other nonlinear electrodynamics which can preserve the conformal invariance of the theory in higher dimensions. It has the form 
\begin{eqnarray}\label{Pmax}
\mathcal{L}(F)=(-F_{\mu\nu}F^{\mu\nu})^s,
\end{eqnarray}
where for the nonlinear parameter $s=1$, it is reduced to the linear Maxwell theory. In order to have an $(n+1)$-dimensional conformal invariant action, the energy-momentum tensor should be traceless which leads to the value, $s=(n+1)/4$. For a general study, we consider an arbitrary value for the parameter $s$. Quasitopological black hole solutions in the presence of the power Maxwell matter field have been obtained in Refs. \cite{power1,power2}. In this section, we aim to investigate the Joule-Thomson expansion of these solutions. The temperture of this black hole follows from \cite{power1,power2}
\begin{eqnarray}\label{Tempmax}
T&=&\frac{1}{4\pi r_{+}(4\mu_{4}k^3+3\mu_{3}k^2r_{+}^2+2\mu_{2}k r_{+}^4+r_{+}^6)}\times [\mu_{4}k^4(n-8)+\mu_{3}k^3(n-6)r_{+}^2+\mu_{2}k^2(n-4)r_{+}^4\nonumber\\
&&+k(n-2)r_{+}^6-\frac{2\Lambda}{n-1} r_{+}^8]-\frac{q^{2s}2^{s}(n-2s)^{2s}r_{+}^{2s(1-n)/(2s-1)}}{4\pi(4\mu_{4}k^3r_{+}^{-6}+3\mu_{3}k^2r_{+}^{-4}+2\mu_{2}k r_{+}^{-2}+1)(n-1)(2s-1)^{2s-1}}r_{+},
\end{eqnarray}
where $r_{+}$ has a rang between $r_{0}\leq r_{+}<\infty$ that $r_{0}\neq0$. Using Eq.\eqref{invesion}, we plot the isenthalpic and inversion curves of the power Maxwell quasitopological solutions in Fig. \ref{Fig6}. It is clear that for a general parameter $s$, the inversion curve has divided the isenthalpic curves into two cooling and heating parts. In the cooling/heating process, the temperature decreases/increases as the pressure decreases in the isenthalpy process. If we compare our results with the Joule-Thomson expansion of the power Maxwell black holes in Einstein gravity \cite{JoulPMax}, they show that the quasitopological gravity cannot make an enhancement for the isenthalpic curves. It is also clear from 
Fig.\ref{Fig6}
 that by increasing the parameter value, $s$, an extreme black hole happens at lower pressure. This is unlike the Einstein-power-Maxwell black hole, in which the extreme black hole with larger $s$ has a larger pressure.\\

\section{Joule-Thomson expansion of the five-dimensional Yang-Mills quasitopological black hole}\label{Joule3}
\begin{figure}
\centering
\subfigure[Yang-Mills quasitopological]{\includegraphics[scale=0.47]{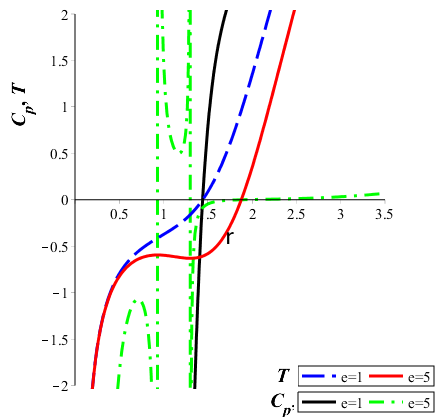}\label{Fig7a}}\hspace*{.2cm}
\subfigure[Maxwell quasitopological]{\includegraphics[scale=0.37]{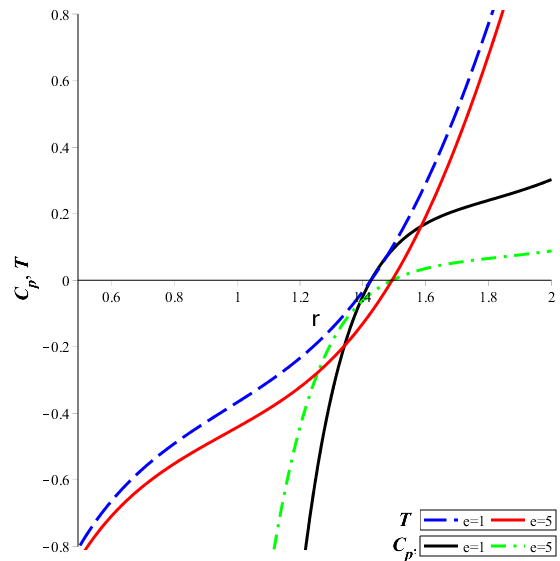}\label{Fig7b}}\caption{ Thermal stability with respect to $r_{+}$ for different values $e$ and $q$ with $\hat{\mu}_{2}=0.1$, $\hat{\mu}_{3}=0.2$, $\hat{\mu}_{4}=0.001$, $k=1$ and $n=4$.}\label{Fig7}
\end{figure}

\begin{figure}
\centering
\subfigure[Yang-Mills quasitopological]{\includegraphics[scale=0.27]{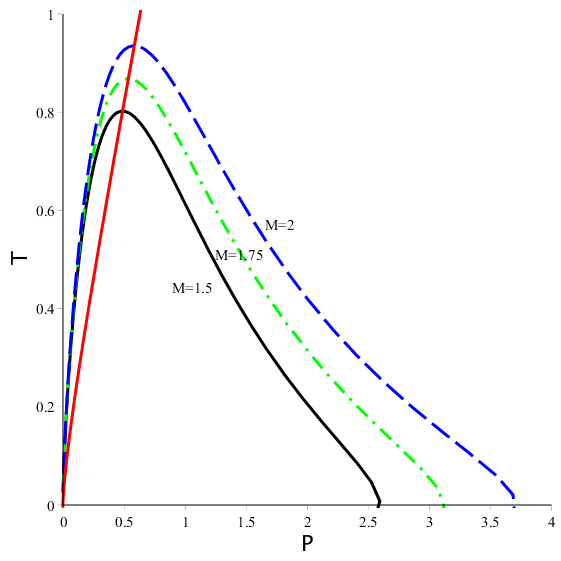}\label{Fig8a}}\hspace*{.2cm}
\subfigure[Maxwell quasitopological]{\includegraphics[scale=0.27]{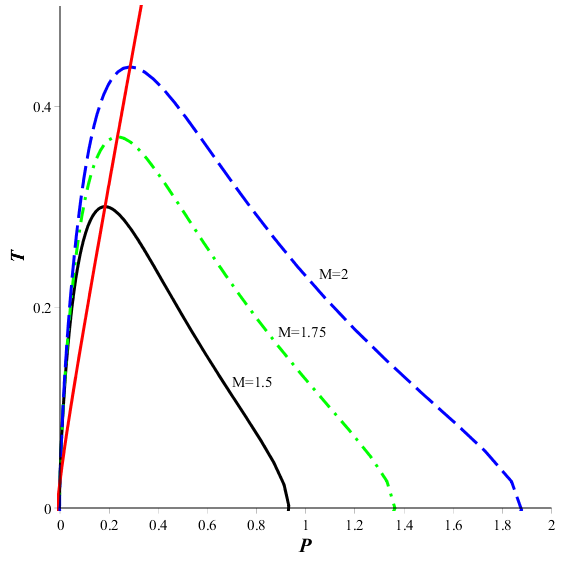}\label{Fig8b}}\caption{Isenthalpic curves and inversion curve with $\hat{\mu}_{2}=0.1$, $\hat{\mu}_{3}=0.2$, $\hat{\mu}_{4}=0.001$, $k=1$, $e=q=5$ and $n=4$.}\label{Fig8}
\end{figure}

In this section, we tend to consider the non-abelian Yang-Mills theory with the quasitopological gravity and obtain the related five-dimensional solutions. We also obtain the thermodynamic quantities of this black hole and then probe the Joule-Thomson expansion for it. In five dimensions, we can just consider the six-parameters gauge groups, $SO(4)$ and $SO(3,1)$, where the action is defined by the relation \eqref{Act1} with the matter source 
\begin{eqnarray}
\mathcal{L}(F)=-\gamma_{ab}F_{\mu\nu}^{(a)}F^{(b)\mu\nu}.
\end{eqnarray}
The gauge field tensor $F_{\mu\nu}$ is described as follows
\begin{eqnarray}
F_{\mu\nu}^{(a)}=\partial_{\mu}A_{\nu}^{(a)}-\partial_{\nu}A_{\mu}^{(a)}+\frac{1}{e}C^{a}_{bc}A_{\mu}^{(b)}A_{\nu}^{(c)}, 
\end{eqnarray}
where $e$ is a coupling constant and $C^{a}_{bc}$'s are the structure constants of the gauge groups that $a,b$ go from 1 to 6. In order to have analytic solutions, we use the Wu-Yang ansatz \cite{Wu} and obtain the gauge potentials of the gauge groups. Using the appropriate coordinates, we have written the gauge potentials and the structure constants of the groups $SO(4)$ and $SO(3,1)$ in the appendix \eqref{app3}. If we vary of the Yang-Mills quasitopological action with respect to $g_{\mu\nu}$, we get to Eq.\eqref{Eq}, where $\xi$ obeys from
\begin{eqnarray}\label{Xi1}
\xi=-\frac{\Lambda}{6}+\frac{m}{r^4}-\frac{2e^2}{r^4}\mathrm{ln}(\frac{r}{r_{0}}),
\end{eqnarray}
and $r_{0}$ is a constant that for simlicity, we choose it, $r_{0}=1$. In order to have the static Yang-Mills quasitopological solutions, we use the metric \eqref{metric1} with $k=-1,+1$ that leads to the solutions \eqref{func1} with the $\xi$ defined in Eq.\eqref{Xi1}. The mass and entropy of the five-dimensional Yang-Mills quasitopological black hole can be gained like Eqs.\eqref{mass} and \eqref{entropy} where $m$ is obtained as bellow
\begin{eqnarray}
m=\mu_{4}\frac{k^4}{r_{+}^4}+\mu_{3}\frac{k^3}{r_{+}^2}+\mu_{2} k^2+k r_{+}^2-\frac{\Lambda}{6}r_{+}^4-2e^2 \mathrm{ln}(r_{+}).
\end{eqnarray}
We can also determine the temperature and the Yang-Mills charge of this black hole as 
\begin{eqnarray}\label{TemYan}
T=-\frac{6\mu_{4}k^4+3\mu_{3}k^3 r_{+}^2-3k r_{+}^6+\Lambda r_{+}^8+3 e^2 r_{+}^4}{6\pi r_{+}(4\mu_{4}k^3+3\mu_{3}k^2r_{+}^2+2\mu_{2} k r_{+}^4+r_{+}^6)},
\end{eqnarray} 
\begin{eqnarray}
Q=\frac{1}{4\pi\sqrt{6}}\int d\Omega_{3} \sqrt{\mathrm{Tr}(F^{(a)}_{\mu\nu}F^{(a)}_{\mu\nu})}=\frac{e}{4\pi}.
\end{eqnarray}
This black hole obeys the first law of thermodynamics
\begin{eqnarray}
dM=TdS+U dQ,
\end{eqnarray}
where $T=\big(\frac{\partial M}{\partial S}\big)_{Q}$ is equal to the temperature \eqref{TemYan} and the Yang-Mills potential $U$ can be obtained by
\begin{eqnarray}
U=\bigg(\frac{\partial M}{\partial Q}\bigg)_{S}=-12 \pi Q \mathrm{ln}(r_{+}).
\end{eqnarray}
This relation restricts the range of the horizon value $r_{+}$ to $1\leq r_{+}<\infty$.
To study the thermal stability of the Yang-Mills quasitopological black hole, we obtain the heat capacity from Eq. \eqref{heatcap} and plot it in Fig.\ref{Fig7a}. We also compare the stability of this black hole with the Maxwell quasitopological black hole in Fig.\ref{Fig7b}. These figures show that for each values of $e$ and $q$, there is a $r_{+\mathrm{min}}$(the condition $r_{+\mathrm{min}}>1$ is established for the Yang-Mills theory) that both $C_{P}$ and $T$ are positive for $r_{+}>r_{+\mathrm{min}}$. For small charges $e$ and $q$, $r_{+\mathrm{min}}$ has a same value in both Yang-Mills and Maxwell quasitopological theories, while for large charges, $r_{+\mathrm{min}}$ has a larger value in the Yang-Mills quasitopological gravity. Therefore, for large $q$, the Maxwell quasitopological black holes have a larger region in thermal stability than the Yang-Mills quasitopological black holes.  
We are also eager to investigate the Joule-Thomson expansion of the five-dimensional Yang-Mills quasitopological black hole in Fig.\ref{Fig8a} and then compare with the Maxwell quasitopological one in Fig.\ref{Fig8b}. In both figures \ref{Fig8a} and \ref{Fig8b}, we can see a cooling/heating process for different values of mass with positive/negative slope in the isenthalpic curves. In fact, as the pressure decreases during this expansion, so according to Eq.\eqref{formule1}, a decrease/increase of the temperture is related to the positive/negative slope in the $T-P$ diagram. For each isenthalpic curve, there is an inversion temperature and pressure, $T_{i}$ and $P_{i}$, which there is a cooling and heating process for $P<P_{i}$ and $P>P_{i}$, respectively. The heating process for Yang-Mills quasitopological black hole in Fig.\ref{Fig8a} happens with a slower slope than the one in Maxwell quasitopological theory. So the extreme Yang-Mills quasitopological black holes are described with larger pressures than the extreme Maxwell quasitopological black holes.    

\section{concluding remarks}\label{con}
In this paper, we reviewed some quasitopological black hole solutions and obtained their thermodynamic properties such as thermal stability and Joule-Thomson expansion for them. At first, we achieved to the $(n+1)$-dimensional static quasitopological black hole solutions in the presence of three BI, EN and LN forms of nonlinear electrodynamics. The obtained solutions are divided into two parts, for $\mu_{4}>0$ and $\mu_{4}<0$. We also obtained the thermodynamic quantities and wrote the first law of thermodynamics in the extended phase space. Then, we sought for the physical existence of the black hole by studying the thermal stability. The stable region of this black hole is independent of the types of the nonlinear electrodynamics. Also for $k=1$, the solutions with $\hat{\mu}_{i}>0$ may lead to a larger stable region than the one with negative $\hat{\mu}_{i}$. Joule-Thomson expansion was our other goal to which we paid attention for this black hole. So, we probed the temperature changes of this black hole in an isenthalpy process during the expansion in which the pressure decreases. For the nonlinear quasitopological black hole with $\hat{\mu}_{4}>0$ and $k=1$, we obtained an inversion curve which can divide the isenthalpic curves in to two parts. The part with the positive slope in isenthalpic curve leads to a cooling process, while for the negative slope one, a heating one may happen. For the black hole with small charge $q$ and nonlinear parameter $\beta$, the temperature reduces to zero with a very slow slope. It is possible to have an isenthalpic curve for $\hat{\mu}_{4}<0$ just for $k=-1$.\\
We also studied the Joule-Thomson expansion of the power Maxwell quasitopological black holes. The results showed that for the large nonlinear parameter, $s$, the extreme black hole has a smaller pressure. This is while that for the Einstein-power-Maxwell black hole, an extreme black hole with a small parameter $s$ happens in small pressure. At the end, we went to the Yang-Mills theory and gained the five-dimensional Yang-Mills solutions in the quasitopological gravity. We also perused the thermodynamic quantities such as thermal stability and Joule-Thomson for this black hole and compared the results with the five-dimensional Maxwell quasitopological black hole. They show that there is a $r_{+\mathrm{min}}$, which the Yang-Mills quasitopological black hole is thermally stable for $r_{+}>r_{+\mathrm{min}}$. For small values of the charges $e$ and $q$, the value of $r_{+\mathrm{min}}$ is independent of the Yang-Mills or Maxwell theories, while, for large charges, $r_{+\mathrm{min}}$ has a larger value in Yang-Mills theory. Also, the heating process for the Yang-Mills quasitopological black hole happens more slowly than the one in the Maxwell black hole.      

\section{Appendix}
\subsection{The coefficients of the quartic quasitopological gravity}\label{app1}
The $c_{i}$'s for the quartic quasitopological theory ${{\mathcal L}_4}$ in Eq. \eqref{quasi4} are defined as
\begin{eqnarray}
&&c_{1}=-(n-1)(n^7-3n^6-29n^5+170n^4-349n^3+348n^2-180n+36)\nonumber\\
&&c_{2}=-4(n-3)(2n^6-20n^5+65n^4-81n^3+13n^2+45n-18)\nonumber\\
&&c_{3}=-64(n-1)(3n^2-8n+3)(n^2-3n+3)\nonumber\\
&&c_{4}=-(n^8-6n^7+12n^6-22n^5+114n^4-345n^3+468n^2-270n+54)\nonumber\\
&&c_{5}=16(n-1)(10n^4-51n^3+93n^2-72n+18)\nonumber\\
&&c_{6}=-32(n-1)^2(n-3)^2(3n^2-8n+3)\nonumber\\
&&c_{7}=64(n-2)(n-1)^2(4n^3-18n^2+27n-9)\nonumber\\
&&c_{8}=-96(n-1)(n-2)(2n^4-7n^3+4n^2+6n-3)\nonumber\\
&&c_{9}=16(n-1)^3(2n^4-26n^3+93n^2-117n+36)\nonumber\\
&&c_{10}=n^5-31n^4+168n^3-360n^2+330n-90\nonumber\\
&&c_{11}=2(6n^6-67n^5+311n^4-742n^3+936n^2-576n+126)\nonumber\\
&&c_{12}=8(7n^5-47n^4+121n^3-141n^2+63n-9)\nonumber\\
&&c_{13}=16n(n-1)(n-2)(n-3)(3n^2-8n+3)\nonumber\\
&&c_{14}=8(n-1)(n^7-4n^6-15n^5+122n^4-287n^3+297n^2-126n+18).\nonumber\\
\end{eqnarray}
\subsection{the definitions for the quasitopological solutions}\label{app2}
\begin{eqnarray}\label{eq1}
A&=&-\frac{3\mu_{3}^2}{8\mu_{4}^2}+\frac{\mu_{2}}{\mu_{4}}\,\,\,\,\,\,\,\,\,\,\,,\,\,\,\,\,\,
B=\frac{\mu_{3}^3}{8\mu_{4}^3}-\frac{\mu_{3}\mu_{2}}{2\mu_{4}^2}+\frac{1}{\mu_{4}}\,\,\,\,\,\,\,,\,\,\,\,\,\, C=-\frac{3\mu_{3}^4}{256 \mu_{4}^4}+\frac{\mu_{2}\mu_{3}^2}{16\mu_{4}^3}-\frac{\mu_{3}}{4\mu_{4}^2}+\frac{\xi}{\mu_{4}},
\end{eqnarray}
\begin{eqnarray}
U=\bigg(-\frac{H}{2}\pm\sqrt{\frac{H^2}{4}+\frac{P^3}{27}}\,\bigg)^{\frac{1}{3}},\,\,P=-\frac{A^2}{12}-C\,\,,\,\,\,\,\,\,\,
H=-\frac{A^3}{108}+\frac{A C}{3}-\frac{B^2}{8},
\end{eqnarray}
\begin{equation}
W=\sqrt{A+2y},\,\,\,\,\,y=\left\{
\begin{array}{ll}
$$-\frac{5}{6}A+U-\frac{P}{3U}$$,\quad \quad\quad\quad \  \ {U\neq 0,}\quad &  \\ \\
$$-\frac{5}{6}A+U-\sqrt[3]{H}$$, \quad \quad\quad\quad{U=0.}\quad &
\end{array}
\right.
\end{equation}
\subsection{the definitions for the Yang-Mills theory}\label{app3}
For this non-abelian gauge theory with N-parameters, we introduce
\begin{eqnarray}\label{gam}
\gamma_{ab}\equiv-\frac{\Gamma_{ab}}{|\mathrm {det} \Gamma_{ab}|^{1/N}},
\end{eqnarray}
where $\Gamma_{ab}=C_{ad}^{c}C_{bc}^{d}$ is the metric tensor of the gauge group and $\mathrm {det} \Gamma_{ab}$ is the determinant of $\Gamma_{ab}$. The indices $a,b,c$ start from 1 and end to $N$. If we introduce the following coordinates 
\begin{eqnarray}
x_{1}&=&\frac{r}{\sqrt{k}}\, \mathrm{sin}(\sqrt{k}\,\theta)\,\mathrm{sin} \phi \,\,\mathrm{sin}\,\psi,\nonumber\\
x_{2}&=&\frac{r}{\sqrt{k}}\, \mathrm{sin}(\sqrt{k}\,\theta)\,\mathrm{sin} \phi \,\,\mathrm{cos}\,\psi,\nonumber\\
x_{3}&=&\frac{r}{\sqrt{k}}\, \mathrm{sin}(\sqrt{k}\,\theta)\,\mathrm{cos} \phi \,\,\nonumber\\
x_{4}&=& r\, \mathrm{cos}\,(\sqrt{k}\,\theta),
\end{eqnarray}
for $SO(4)$ gauge group ($k=1$ and $n=5$)
\begin{eqnarray}
C^{1}_{24}&=& C^{1}_{35}=C^{2}_{41}=C^{2}_{36}=C^{3}_{51}=C^{3}_{62}=1,\nonumber\\
C^{4}_{56}&=& -C^{4}_{21}=C^{5}_{64}=-C^{5}_{31}=C^{6}_{45}=-C^{6}_{32}=1,\nonumber\\
\gamma_{ab}&=&\mathrm{diag}(1,1,1,1,1,1),
\end{eqnarray}
\begin{eqnarray}
A_{\mu}^{(1)}&=& -e\,(\mathrm{sin}\, \phi\,\mathrm{cos}\, \psi\, d\theta+\mathrm{sin}\, \theta \,\mathrm{cos}\,\theta \,(\mathrm{cos}\,\phi\, \mathrm{cos}\,\psi \,d\phi-\mathrm{sin}\,\phi\, \mathrm{sin}\,\psi \,d\psi))\nonumber\\
A_{\mu}^{(2)}&=& -e\,(\mathrm{sin}\, \phi\,\mathrm{sin}\, \psi\, d\theta+\mathrm{sin}\, \theta \,\mathrm{cos}\,\theta \,(\mathrm{cos}\,\phi\, \mathrm{sin}\,\psi \,d\phi+\mathrm{sin}\,\phi\, \mathrm{cos}\,\psi \,d\psi))\nonumber\\
A_{\mu}^{(3)}&=& -e\,(\mathrm{cos}\, \phi\, d\theta-\mathrm{sin}\, \theta \,\mathrm{cos}\,\theta \,\mathrm{sin}\,\phi \,d\phi)\nonumber\\
A_{\mu}^{(4)}&=& -e\,\mathrm{sin}^{2}\, \theta \,\mathrm{sin}^{2}\, \phi \,d\psi\nonumber\\
A_{\mu}^{(5)}&=& e\,\mathrm{sin}^{2}\, \theta\,(\mathrm{cos}\, \psi\, d\phi-\mathrm{sin}\, \phi \,\mathrm{cos}\,\phi \,\mathrm{sin}\,\psi \,d\psi)\\
A_{\mu}^{(6)}&=& e\,\mathrm{sin}^{2}\, \theta\,(\mathrm{sin}\, \psi\, d\phi+\mathrm{sin}\, \phi \,\mathrm{cos}\,\phi \,\mathrm{cos}\,\psi \,d\psi)
\end{eqnarray}
for $SO(3,1)$ gauge group ($k=-1$ and $n=5$)
\begin{eqnarray}
C^{1}_{24}&=& C^{1}_{35}=C^{2}_{41}=C^{2}_{36}=C^{3}_{51}=C^{3}_{62}=1\nonumber\\
C^{4}_{56}&=& C^{4}_{21}=C^{5}_{64}=C^{5}_{31}=C^{6}_{45}=C^{6}_{32}=1\nonumber\\
\gamma_{ab}&=&\mathrm{diag}(-1,-1,-1,1,1,1),
\end{eqnarray}
\begin{eqnarray}
A_{\mu}^{(1)}&=& -e\,(\mathrm{sin}\, \phi\,\mathrm{cos}\, \psi\, d\theta+\mathrm{sinh}\, \theta \,\mathrm{cosh}\,\theta \,(\mathrm{cos}\,\phi\, \mathrm{cos}\,\psi \,d\phi-\mathrm{sin}\,\phi\, \mathrm{sin}\,\psi \,d\psi)),\nonumber\\
A_{\mu}^{(2)}&=& -e\,(\mathrm{sin}\, \phi\,\mathrm{sin}\, \psi\, d\theta+\mathrm{sinh}\, \theta \,\mathrm{cosh}\,\theta \,(\mathrm{cos}\,\phi\, \mathrm{sin}\,\psi \,d\phi+\mathrm{sin}\,\phi\, \mathrm{cos}\,\psi \,d\psi)),\nonumber\\
A_{\mu}^{(3)}&=& -e\,(\mathrm{cos}\, \phi\, d\theta-\mathrm{sinh}\, \theta \,\mathrm{cosh}\,\theta \,\mathrm{sin}\,\phi \,d\phi),\nonumber\\
A_{\mu}^{(4)}&=& e\,\mathrm{sinh}^{2}\, \theta \,\mathrm{sin}^{2}\, \phi \,d\psi,\nonumber\\
A_{\mu}^{(5)}&=& -e\,\mathrm{sinh}^{2}\, \theta\,(\mathrm{cos}\, \psi\, d\phi-\mathrm{sin}\, \phi \,\mathrm{cos}\,\phi \,\mathrm{sin}\,\psi \,d\psi),\\
A_{\mu}^{(6)}&=& -e\,\mathrm{sinh}^{2}\, \theta\,(\mathrm{sin}\, \psi\, d\phi+\mathrm{sin}\, \phi \,\mathrm{cos}\,\phi \,\mathrm{cos}\,\psi \,d\psi).
\end{eqnarray}
\acknowledgments{This work is supported by Irainian National Science Foundation (INSF). F.N would like to thank physics department of Isfahan University of Technology for warm hospitality.}


\begin{thebibliography}{99}

\bibitem{2}
 O. Aharony, S. S. Gubser, J. M. Maldacena, H. Ooguri, and Y. Oz, {\emph {``Large N Field Theories, String Theory and Gravity"}}, Phys. Rep. \textbf{323}, 183 (2000). \hyperref{https://arxiv.org/abs/hep-th/9905111}{}{}{arxiv:hep-th/9905111}.
 
 \bibitem{4}
 S. W. Hawking and D. N. Page, {\emph {``Thermodynamics of Black Holes in anti-De Sitter Space"}},  Commun. Math. Phys. \textbf{87}, 577 (1983). 
 


\bibitem{30}
 A. Chamblin, R. Emparan, C. V. Johnson and R. C. Myers, {``\emph {Charged AdS Black Holes and Catastrophic Holography"}},  Phys. Rev. D \textbf{60}, 064018 (1999). \hyperref{https://arxiv.org/abs/hep-th/9902170}{}{}{arxiv:hep-th/9902170}.
 
 \bibitem{32}
 A. Chamblin, R. Emparan, C. V. Johnson and R. C. Myers, {\emph {``Holography, thermodynamics, and fluctuations of charged AdS black holes"}},  Phys. Rev. D \textbf{60}, 104026 (1999). \hyperref{https://arxiv.org/abs/hep-th/9904197}{}{}{arxiv:hep-th/9904197}.

\bibitem{115}
D. Kastor, S. Ray and J. Traschen, {\emph {``Enthalpy and the Mechanics of AdS Black Holes"}}, Class. Quant. Grav. \textbf{26}, 195011 (2009).\hyperref{https://arxiv.org/abs/0904.2765}{}{}{arxiv:0904.2765 [hep-th]}.

\bibitem{okcu1}
$\ddot{O}$. $\ddot{O}$kc$\ddot{u}$ and E. Aydiner, {\emph {``Joule-Thomson Expansion of Charged AdS Black Holes"}}, Eur. Phys. J. C \textbf{77}, 24 (2017); \hyperref{https://arxiv.org/abs/1611.06327}{}{}{arXiv:1611.06327 [gr-qc]}.

\bibitem{39}
 J. X. Mo, G. Q. Li, S. Q. Lan and X. B. Xu, {\emph {``Joule-Thomson expansion
 		of d-dimensional charged AdS black holes"}},  Phys. Rev. D \textbf{98}, 124032  (2018). \hyperref{https://arxiv.org/abs/1804.02650}{}{}{arXiv:1804.02650 [gr-qc]}.
 
\bibitem{40}
 S. Q. Lan, {\emph {``Joule-Thomson expansion of charged Gauss-Bonnet black holes in AdS space"}}, Phys. Rev. D \textbf{98}, 084014 (2018).\hyperref{https://arxiv.org/abs/1805.05817}{}{}{arXiv:1805.05817 [gr-qc]}.

\bibitem{42}
 H. Ghaffarnejad, E. Yaraie and M. Farsam, {\emph {``Quintessence Reissner Nordström Anti de Sitter Black Holes and Joule Thomson Effect"}}, Int. J. Theor. Phys. \textbf{57}, 1671 (2018).\hyperref{https://arxiv.org/abs/1802.08749}{}{}{arXiv:1802.08749 [gr-qc]}.
 
\bibitem{43Almeida}
 R. D. Almeida and K. P. Yogendran, {\emph {``Thermodynamic Properties of Holographic superfluids"}}, \hyperref{https://arxiv.org/abs/1802.05116}{}{}{arXiv:1802.05116}. 
 
\bibitem{44}
   C. L. Ahmed Rizwan, A. Naveena Kumara, Deepak Vaid and K. M. Ajith, {\emph {``Joule-Thomson expansion in AdS black hole with a global monopole"}}. \hyperref{https://arxiv.org/abs/1805.11053}{}{}{arXiv:1805.11053}.
 

\bibitem{41}
 M. Chabab, H. El Moumni, S. Iraoui, K. Masmar and S. Zhizeh, {\emph {``Joule-Thomson Expansion of RN-AdS Black Holes in f(R) gravity"}}, JHEP \textbf{02}, 05  (2018).\hyperref{https://arxiv.org/abs/1804.10042}{}{}{	arXiv:1804.10042 [gr-qc]}.
\bibitem{Rostami} M. Rostami, J. Sadeghi, S. Miraboutalebi, A. A. Masoudi and B. Pourhassan, {\emph {``Charged accelerating AdS black hole of f(R) gravity and the Joule-Thomson expansion"}}.\hyperref{https://arxiv.org/abs/1908.08410}{}{} {arXiv:1908.08410}.

\bibitem{46} J.X. Mo and G.Q. Li, {\emph {``Effects of Lovelock gravity on the Joule-Thomson expansion"}}.\hyperref{https://arxiv.org/abs/1805.04327}{}{} {arXiv:1805.04327}. 

\bibitem{Jrainbow} D. M. Yekta, A. Hadikhani and $\ddot{O}$. $\ddot{O}$kc$\ddot{u}$, {\emph {``Joule-Thomson expansion of charged AdS black holes in rainbow	gravity"}}, Phys. Lett. B, \textbf{795}, 521. (2019). \hyperref{https://arxiv.org/abs/1905.03057}{}{}{arXiv:1905.03057 [hep-th]}.
  \bibitem{Guo}
  J. Pu, S. Guo, Q. Q. Jiang and X. T. Zu, {\emph {``Joule-Thomson expansion of the regular(Bardeen)-AdS black hole"}}, Chin. Phys. C {\bf 44}, 035102 (2020).\hyperref{https://arxiv.org/abs/1905.02318}{}{}{	arXiv:1905.02318 [gr-qc]}.

\bibitem{Li}
 C. Li, P. He, P. Li and J. B. Deng, {\emph {``Joule-Thomson expansion of the Bardeen-AdS black holes"}}, Gen. Rel. Grav. \textbf{52}, 50 (2020).\hyperref{https://arxiv.org/abs/1904.09548}{}{}{arXiv:1904.09548 [gr-qc]}.
   
\bibitem{JoulBI} S. Bi, M. Du, J., Tao and F. Yao, {\emph {``Joule-Thomson expansion of Born-Infeld AdS black holes"}}, Chinese Physics C  \textbf{45} (2), 025109 (2020); \hyperref{https://arxiv.org/abs/2006.08920}{}{}{arXiv:2006.08920 [gr-qc]}.
\bibitem{Born} M. Born and L. Infeld, {\emph {``Foundations of the new field theory"}}, Proc. R. Soc. A \textbf{144}, 425 (1934).
\bibitem{Soleng} H. H. Soleng, {\emph {``Charged black points in general relativity coupled to the logarithmic U(1) gauge theory"}}, Phys. Rev. D \textbf{52}, 6178 (1995).
\bibitem{Hen00} S. H. Hendi, {\emph {``Asymptotic charged BTZ black hole solutions"}}, J. High Energy Phys. \textbf{03}, 065 (2012); 	arXiv:1405.4941 [hep-th].\hyperref{https://arxiv.org/abs/1405.4941}{}{}{arXiv:1405.4941 [hep-th]}.
\bibitem{Lemos}J. P. Lemos, {\emph {``Cylindrical Black Hole in General Relativity"}}, Classical Quantum Gravity \textbf{12}, 1081 (1995); Phys. Lett. B \textbf{353}, 46 (1995);\hyperref{https://arxiv.org/abs/gr-qc/9404041}{}{}{arXiv:gr-qc/9404041}.
\bibitem{Cai} R. G. Cai and Y. Z. Zhang, {\emph {``Black plane solutions in four-dimensional spacetimes"}}, Phys. Rev. D \textbf{54}, 4891 (1996).
\bibitem{Mann} R. B. Mann, {\emph {``Pair production of topological anti-de Sitter black holes"}}, Classical Quantum Gravity \textbf{14}, L109 (1997).
\bibitem{Myers} R. C. Myers, M. F. Paulos, and A. Sinha, {\emph {``Holographic studies of quasi-topological gravity"}}, J. High Energy Phys. \textbf{08}, 035 (2010). \hyperref{https://arxiv.org/abs/1004.2055}{}{}{arXiv:1004.2055 [hep-th]}.
     
\bibitem{Mann2012}  Dehghani, M.H., Bazrafshan, A., Mann, R.B., Mehdizadeh, M.R., Ghanaatian, M. and Vahidinia,, M.H., {\emph {``Black holes in (quartic) quasitopological gravity"}}, Phys. Rev. D, \textbf{85} (10), 104009 (2012); \hyperref{https://arxiv.org/abs/1109.4708}{}{}{arXiv:1109.4708 [hep-th]}.
\bibitem{Man1} R. A. Hennigar, W. G. Brenna and R. B. Mann,{\emph {``P-v criticality in quasitopological gravity"}}, JHEP \textbf{1507}, 077 (2015).\hyperref{https://arxiv.org/abs/1505.05517}{}{}{arXiv:1505.05517 [hep-th]}.
\bibitem{Mir1} M. Mir, R. A. Hennigar, J. Ahmed, and R. B. Mann,{\emph {``Black hole chemistry and holography in generalized quasi-topological gravity"}}, Journal of High Energy Physics, \textbf{08}, 068 (2019). \hyperref{https://arxiv.org/abs/1902.02005}{}{}{arXiv:1902.02005 [hep-th]}.
\bibitem{Mir2}  M. Mir and R. B. Mann, {\emph {``On generalized quasi-topological cubic-quartic gravity: thermodynamics and holography"}}, Journal of High Energy Physics 2019, no. \textbf{7} (2019).\hyperref{https://arxiv.org/abs/1902.10906}{}{}{arXiv:1902.10906 [hep-th]}.
\bibitem{yass1} P.B. Yasskin, {\emph{``Solutions for gravity coupled to massless gauge fields"
}}, Phys. Rev. D \textbf{12}, 2212 (1975). 
\bibitem{18ym} G. Lavrelashvili and D. Maison,\emph{``Regular and black hole solutions of Einstein-Yang-Mills dilaton theory
''}, Nucl. Phys. B \textbf{410}, 407 (1993). \hyperref{https://www.sciencedirect.com/science/article/pii/055032139390441Q}{}{}{}.

\bibitem{19ym} E.E. Donets and D.V. Galtsov, \emph{``Stringy Sphalerons and Non-Abelian Black Holes''}, Phys. Lett. B \textbf{302}, 411 (1993), \hyperref{https://arxiv.org/abs/hep-th/9212153}{}{}{arXiv:hep-th/9212153}.

\bibitem{21ym} T. Torii  and K. Maeda,\emph{``Black holes with non-Abelian hair and their thermodynamical properties
''}, Phys. Rev. D \textbf{48}, 1643 (1993).\hyperref{https://journals.aps.org/prd/abstract/10.1103/PhysRevD.48.1643}{}{}{}.

\bibitem{22ym} Y. Brihaye and E. Radu,\emph{``Euclidean solutions in Einstein-Yang-Mills-dilaton theory
''}, Phys. Lett. B {\textbf 636}, 212 (2006).\hyperref{https://arxiv.org/abs/gr-qc/0602069}{}{}{arXiv:gr-qc/0602069}.

\bibitem{6ym} M. Wirschins, A. Sood, J. Kunz,{\emph {``Non-Abelian Einstein-Born-Infeld black holes"}}, Phys. Rev. D \textbf{63}, 084002 (2001).\hyperref{https://arxiv.org/abs/hep-th/0004130}{}{}{arXiv:hep-th/0004130}.

\bibitem{7ym} M. Huebscher, P. Meessen, T. Ortin, S. Vaula, \emph{``N=2 Einstein-Yang-Mills's BPS solutions
''}, J. High Energy Phys.\textbf{09}, 099 (2008).\hyperref{https://arxiv.org/abs/0806.1477}{}{}{arXiv:0806.1477 [gr-qc]}.


\bibitem{Wu} T. T. Wu, C. N. Yang {\emph {in Properties of Matter Under Unusual Conditions, edited by }} H. Mark, New York, London: Interscience, 349 (1969).
\bibitem{2ym} S.H. Mazharimousavi and M. Halilsoy, \emph{``5D-Black Hole Solution in Einstein-Yang-Mills-Gauss-Bonnet Theory
''}, Phys. Rev. D \textbf{76}, 087501 (2007),\hyperref{https://arxiv.org/abs/0801.1562}{}{}{arXiv:0801.1562 [gr-qc]}.

\bibitem{3ym} S.H. Mazharimousavi, M. Halilsoy,\emph{``Black Hole solutions in Einstein-Maxwell-Yang-Mills-Gauss-Bonnet Theory
''} J. Cosmol. Astropart. Phys. \textbf{12}, 005 (2008). \hyperref{https://arxiv.org/abs/0801.2110}{}{}{arXiv:0801.2110 [gr-qc]}.

\bibitem{4ym} S.H. Mazharimousavi and M. Halilsoy, \emph{``Higher dimensional Yang-Mills black holes in third order Lovelock gravity
''},Phys. Lett. B \textbf{665}, 125 (2008). \hyperref{https://arxiv.org/abs/0801.1726}{}{}{arXiv:0801.1726 [gr-qc]}.

\bibitem{5ym} S.H. Mazharimousavi and M. Halilsoy,\emph{``Lovelock black holes with a power-Yang-Mills source
''}, Phys. Lett. B \textbf{681}, 190 (2009). \hyperref{https://arxiv.org/abs/0908.0308}{}{}{arXiv:0908.0308 [gr-qc]}.

\bibitem{Bost} N. Bostani and M. H. Dehghani,\emph{``Topological Black Holes of (n+1)-dimensional Einstein-Yang-Mills Gravity
''}, Mod. Phys. Lett. A \textbf{25}, 1507 (2010). \hyperref{https://arxiv.org/abs/0908.0661}{}{}{
 arXiv:0908.0661 [gr-qc]}.

\bibitem{Deh} M. H. Dehghani, N. Bostani and R. Pourhasan,\emph{``Topological black holes of Gauss-Bonnet-Yang-Mills gravity''}, Int. J. Mod. Phys. D, \textbf{19}, 1107 (2010). \hyperref{https://arxiv.org/abs/0908.0663}{}{}{ arXiv:0908.0663 [gr-qc]}.

\bibitem{24ym} M.H. Dehghani, A. Bazrafshan, {\emph {``Topological black holes of Einstein-Yang-Mills dilaton gravity"}}, Int. J. Mod. Phys. D \textbf{19},  293 (2010).\hyperref{https://arxiv.org/abs/1005.2387}{}{}{arXiv:1005.2387 [gr-qc]} . 

\bibitem{Baz1} A. Bazrafshan, F. Naeimipour, M. Ghanaatian and A. Khajeh,{\emph {``Physical and thermodynamic properties of quartic quasitopological black holes and rotating black branes with a nonlinear source"}},  Phys. Rev. D \textbf{100}, 064062 (2019).\hyperref{https://arxiv.org/abs/1905.12428}{}{}{arXiv:1905.12428 [gr-qc]}.  
\bibitem{Ghana1} M. Ghanaatian,\emph{``Quartic quasi-topological-Born–Infeld gravity
''}, General Relativity and Gravitation, \textbf{47}, 105 (2015).\hyperref{https://arxiv.org/abs/1503.09053}{}{}{arXiv:1503.09053}.

\bibitem{Naeimi3} M. Ghanaatian, F. Naeimipour, A. Bazrafshan and M. Abkar,\emph{``Charged black holes in quartic quasi-topological gravity''} Phys. Rev. D \textbf{97}, 104054 (2018).\hyperref{https://arxiv.org/abs/1801.05692}{}{}{arXiv:1801.05692 [gr-qc]}.



\bibitem{Brown} J.D. Brown and J. W. York Jr, \emph{``Quasilocal energy and conserved charges derived from the gravitational action''}, Phys. Rev. D {\bf 47} 1407 (1993). \hyperref{https://arxiv.org/abs/gr-qc/9209012}{}{}{arXiv:gr-qc/9209012}.


\bibitem{Wald} R. M. Wald,\emph{``Black hole entropy is the Noether charge''}, Phys. Rev. D \textbf{48}, 3427 (1993). \hyperref{https://arxiv.org/abs/gr-qc/9307038}{}{}{arXiv:gr-qc/9307038}.

\bibitem{Dolan} B. P. Dolan, \emph {``Pressure and volume in the first law of black hole thermodynamics''}, Class. Quant. Grav. \textbf{28} 235017 (2011). \hyperref{https://arxiv.org/abs/1106.6260}{}{}{arXiv:1106.6260 [gr-qc]}.

\bibitem{power1} M. Ghanaatian, F. Naeimipour, A. Bazrafshan, M. Eftekharian and A. Ahmadi,{\emph {``Third order quasitopological black hole with a power-law Maxwell nonlinear source"}}, Phys. Rev. D \textbf{99}, 024006 (2019). \hyperref{https://arxiv.org/abs/1809.05198}{}{}{arXiv:1809.05198 [gr-qc]}. 

\bibitem{power2} M. Ghanaatian and A. Bazrafshan,{\emph {``Nonlinear Charged Black Holes in AdS Quasi-Topological Gravity"}}, Int. J. Mod. Phys. D \textbf{22}, 1350076 (2013). \hyperref{https://arxiv.org/abs/1304.2311}{}{}{arXiv:1304.2311 [gr-qc]}.

\bibitem{JoulPMax} Z. W. Feng, X. Zhou and S. Q. Zhou,{\emph {``Joule-Thomson expansion of higher dimensional nonlinearly charged AdS black hole in Einstein-PMI gravity"}}.\hyperref{https://arxiv.org/abs/2009.02172}{}{}{arXiv:2009.02172 [gr-qc]}.  

\end{thebibliography}
\end{document}